\documentstyle[prl,aps,twocolumn]{revtex}
\large  

\input epsf
\begin{document}
\draft \twocolumn[\hsize\textwidth
\columnwidth\hsize\csname
@twocolumnfalse\endcsname

\title{Self-localization of directed polymers and oppressive population 
control} 
\author{T. J. Newman\cite{tea} and Eugene B. Kolomeisky\cite{eea}} 
\address{Department of Physics, University of Virginia,
McCormick Road, Charlottesville, VA 22903} 
\maketitle
\begin{abstract}
We construct a phenomenological theory of self-localization of directed
polymers in $d+1$ dimensions. In $d=1$ we show that the polymer is 
always self-localized, whereas in $d=2$ there is a phase
transition between localized and free states.
We also map this system to a model of population dynamics
with fixed total population. Our previous results translate to
static and expanding population clusters, depending on the
birth and death rates. A novel ``pseudo-travelling wave'' is
observed in some sectors of parameter space. 
\end{abstract}
\vspace{2mm} 
]

\narrowtext

Directed polymers are important topological objects in many
condensed matter systems. Examples are flux lines in superconductors\cite{sc},
domain walls in ferromagnets\cite{mag}, and atomic steps on 
vicinal surfaces\cite{terr}. In the case of an isolated directed polymer, 
theoretical studies have concentrated on the effects of quenched
disorder, which may be either point-like or columnar. Using
disorder to suppress the wandering of flux lines is a central
concept in designing useful high-temperature superconductors\cite{super}.

Another mechanism by which the wandering of directed polymers
may be suppressed is self-localization, mediated by elastic
distortions of the medium in which the polymer is embedded.
This has been studied previously via a variational
free energy functional\cite{ej}, which is the analogue 
of the quantum action for the polaron problem\cite{rash}. In the
first part of this Letter we study the phenomenon from an alternative
viewpoint, using the statistical mechanics of directed polymers.
Such an approach leads us to a simple phenomenological equation
for the probability density $P$ of the directed polymer. 
We show that in $d=1$ the
directed polymer is always self-localized, while in $d=2$ the
polymer undergoes a transition to self-localization above
a critical value of the coupling constant $\lambda$ (which
measures the strength of interaction between the polymer
and the embedding environment). The details of this
transition are investigated by numerical integration.

In the latter half of this Letter we show that the phenomenological 
equation for $P$ may be re-interpreted in the language of 
population dynamics. In this case $P$ corresponds to the
local population density. The precise form of the dynamics
is unusual, since along with the birth and death terms,
we have a global constraint that fixes the total population.
[Similar constraints have been studied recently in the context of 
plant population models\cite{gc}, and a link has been made to
self-organized criticality. Also, interesting links have been made between
``non-Hermitian quantum mechanics'' and population biology\cite{nel}.]
The issue in the present work is not whether the population
thrives or becomes extinct, but rather how the individuals
spatially distribute themselves. Different
phases are observed corresponding to self-localized and
expanding populations. We also test the predictions of the
theory by simulations of a discrete reaction-diffusion model
with a global constraint. Finally, within the context of
population dynamics we are at liberty to reverse the sign
of the coupling constant $\lambda $. In this case, we find
that in $d=1$ the density is delocalized via
a pseudo-travelling wave -- an unusual scenario which interpolates 
between dynamical scaling and travelling waves. We end the Letter
by noting the general mapping between directed polymers
and constrained population dynamics.

We consider a directed polymer with elastic constant $\kappa $
in a space with $d$ transverse
dimensions ${\bf r}$, and one longitudinal dimension $t$.
The polymer experiences thermal fluctuations from a
reservoir at temperature $T$, and a potential $V({\bf r},t)$.
One can write the partition function $Z({\bf r},t)$ 
for such a polymer with ends fixed at $({\bf 0},0)$ and $({\bf r},t)$
in the form of a path integral. Following standard methods\cite{schul}, this
path integral may be rewritten as a partial differential
equation of the form
\begin{equation}
\label{pdez}
\partial_{t} Z = \nu \nabla ^{2} Z - V Z \ ,
\end{equation}
where $\nu = T/2\kappa$ and we have absorbed a factor of $T$ into the
potential. The probability density of the polymer is constructed
by normalizing $Z$: $P({\bf r},t) = Z({\bf r},t)
/ \int d^{d}r' \ Z({\bf r}',t)$. Inserting this definition into
the equation for $Z$ yields a closed equation for $P$:
\begin{equation}
\label{pdep}
\partial_{t} P = \nu \nabla ^{2} P - V P
+ \left [ \int d^{d}r' V({\bf r}',t) P({\bf r}',t) \right ] P \ .
\end{equation}   
To our knowledge, this description of the directed polymer has not 
been studied previously. Note this equation does not have the form
of a Fokker-Planck equation -- conservation of probability is
not achieved via an equation of continuity. The physics of
directed polymers is intrinsically non-local.

We want to consider the potential $V$ to have been
caused purely from elastic distortions of the embedding environment,
due to the presence of the directed polymer itself. In principle
we could formulate a detailed microscopic model which includes these
elastic degrees of freedom explicitly, and then attempt to integrate
them out to find the effective potential. However, we are content
to take a phenomenological viewpoint: the elastic
distortions caused by the polymer will provide an attractive 
potential which is strongest in the immediate vicinity of the
polymer, and negligible far from the polymer. If we try to 
construct such a potential from the probability density $P$, we 
come at once upon the particularly simple form $V({\bf r},t)
= -\lambda P({\bf r},t)$, where $\lambda$ is a non-negative coupling
constant. Inserting this relation
into Eq.(\ref{pdep}) we have a closed equation for $P$ of the 
form
\begin{equation}
\label{scpdep}
\partial_{t} P = \nu \nabla ^{2} P + \lambda P^{2} - \lambda \sigma (t)P \ ,
\end{equation}
where $\sigma (t) \equiv \int d^{d}r \ P({\bf r},t)^{2}$.

Given we have a non-linear integro-differential equation for
$P$ there is little hope of a complete analytic solution. However
we will uncover much of the physics from scaling arguments, 
and numerical integration. First, let us consider the possibility
of simple scaling, and make a similarity Ansatz, which, based
on the radial symmetry of the problem, we take as
$P({\bf r},t) = \xi (t)^{-d}F(z)$, with $z=r/\xi$. Note, the prefactor
is fixed by normalization of $P$. Inserting this Ansatz into
Eq.(\ref{scpdep}) yields
\begin{eqnarray}
\label{scal1}
\nonumber
{\dot \xi}(dF+zF')+(\nu/\xi)(F''+(d-1)F'/z) & + & \\
(\lambda/\xi ^{d-1}) (F-\sigma _{0})F & = & 0 \ ,
\end{eqnarray}
where $\sigma _{0} = S_{d}\int dz z^{d-1} F(z)^{2}$, and $S_{d}$ is
the area of the unit hypersphere. We need to balance
the terms arising from the derivative with respect to $t$ and the Laplacian.
This implies ${\dot \xi} = (\nu/\xi)$ which in turn gives
$\xi = (2\nu t)^{1/2}$. So for large $t$ the correlation
length $\xi $ is large, and we see on comparing the Laplacian
terms to the interaction terms (i.e. those proportional to 
$\lambda $) that for $d>2$ the interaction terms are negligible.
In other words, for $d>2$ the behaviour of $P$ corresponds to
thermal diffusion (at least for modest values of $\lambda $).

With the assumption of a $t$-dependent correlation length, we
see that the interaction terms dominate for $d<2$, and there
is no self-consistent solution. We therefore demand
that $\xi $ is $t$-independent, and denote it as $\xi _{0}$.
Focusing on $d=1$, we find
\begin{equation}
\label{stat1}
F''+ (\lambda \xi _{0}/\nu) (F-\sigma _{0})F = 0 \ .
\end{equation}
This equation is simple enough to yield an exact solution.
Expressed in terms of the original spatial variable $x$,
we find the $t$-independent solution
\begin{equation}
\label{stat2}
P(x) = (1/4\xi_{0}) {\rm sech}^{2} (x/2\xi_{0}) \ ,
\end{equation}
where the localization length is self-consistently found
to be $\xi _{0} = 6\nu/\lambda $. We have confirmed the 
stability of this solution by direct numerical integration
of Eq.(\ref{scpdep}) using a variety of initial conditions. In 
all cases we find rapid convergence to the form given above.
It is remarkable that our solution coincides exactly with
the well-known form for the polaron probability density, first found
by Rashba\cite{rash}. This agreement is satisfying from
a physical viewpoint, but by no means obvious given
the mathematical difference between Eq.(\ref{scpdep}) and
the non-linear Schr\"odinger equation governing polaron
physics.

We can achieve a perfect balance of all the terms in Eq.(\ref{scal1})
in exactly two dimensions. In this case, the correlation length drops
out of the equation and we have an ordinary differential
equation for $F(z)$ of the form
\begin{equation}
\label{scal2}
F''+(z+1/z)F'+2F+{\tilde \lambda}(F-\sigma _{0})F = 0 \ ,
\end{equation}
where ${\tilde \lambda }=\lambda /\nu$ is a dimensionless
coupling constant. We have been unable to derive an
analytic solution to this equation. We have studied
its properties using a straightforward numerical integration
scheme, starting with a localized initial condition,
generally taken to be Gaussian. For modest values of ${\tilde \lambda }$
the probability density diffuses outwards,
and we find good data collapse in accordance with the
similarity Ansatz defined in terms of $z=r/\sqrt{t}$. 
As we increase ${\tilde \lambda }$ we
find a sudden transition at ${\tilde \lambda }_{c} \simeq 30.5$,
above which the probability density shrinks to a microscopically
collapsed state. Details of this transition will be
given in a longer paper\cite{kn}.

We now move away from the physics of directed polymers, and
re-interpret our central equation (\ref{scpdep}) within a
very different context, namely reaction kinetics or 
population dynamics. Consider a species $A$ (either chemical,
biological or {\it homo sapiens}) which undergoes the following birth and
death processes: $A \rightarrow 0$ and 
$2A \rightarrow 3A$, with rates $k$ and $k'$ respectively, which
are in general time dependent. We denote the relative density
of $A$ by $\rho ({\bf x},t)$, and stress that the symbol $t$ now
represents real time. We may write a
reaction-diffusion equation to describe these processes:
\begin{equation}
\label{rde}
\partial_{t} \rho = D \nabla ^{2} \rho + k'(t)\rho ^{2} - 
k(t)\rho \ ,
\end{equation}
where $D$ is the effective diffusion constant of individual
$A$ `particles'. The above equation is only an approximate
description of this process, and {\it a priori} is only
expected to be valid above some critical dimension $d_{u}$.
For lower dimensions, it is often necessary
to explicitly include the appropriate (microscopically derived)
noise term in the above equation\cite{mnoise}. For 
a qualitative understanding of the model (which is further refined below), 
such a noise term is not required.

At this point, we shall introduce a somewhat
unusual feature -- a global constraint on the population.
In other words we fix the normalization of the relative density
to unity: $\int d^{d}r \rho = 1$. From Eq.(\ref{rde}) we see this imposes 
$k'(t)/k(t) = \int d^{d}r \rho ^{2}$.
We can imagine two special cases of the above. First, we can
fix the death rate $k$ to be constant, which implies that the
total population is controlled by tuning the
birth rate $k'$ (which is not unknown).
Alternatively, we have the chilling mechanism of total population control
in which the birth rate is fixed, equal to $k_{0}$ say, 
and the death rate is adjusted according to 
$k(t) = k_{0}\int d^{d}x \rho ^{2}$. 
At the level of our phenomenological description,
these two choices lead to similar behaviour, so we shall
concentrate on the second. Substituting the appropriately
tuned rates into Eq.(\ref{rde}) we have
\begin{equation}
\label{rde2}
\partial_{t} \rho = D \nabla ^{2} \rho + k_{0}\rho ^{2} - 
k_{0}\left [ \int d^{d}r' \rho ({\bf r}',t)^{2} \right ] 
\rho \ .
\end{equation}
This equation is identical in form to the self-consistent
equation (\ref{scpdep}) for the probability density of the directed
polymer. Thus all of our results may be taken over and
immediately applied to the population system outlined
above. We state the gross features here. In $d=1$ the 
population will distribute itself in a stationary localized
structure, with a spatial scale $\xi _{0} \sim D/k_{0}$. In $d=2$
the population will attempt to spread itself as widely as possible,
with a diffusive dynamics, but only for $k_{0}$ less than 
some critical value $k_{c}$. Above this value, the system
will undergo collapse and exist in a stationary 
high-density phase. For $d=3$ the population will generally
exist in the diffusive spreading phase (although there is
a possibility for different behaviour for very high values
of $k_{0}$). 

As mentioned before, it is not obvious {\it a priori} whether
a phenomenological description such as Eq.(\ref{rde2}) is an
accurate model of the true population dynamics for low
dimensions. Intuitively, we expect the microscopic fluctuations to
be suppressed in our model due to the long-range interactions
which are built-in via the global population constraint.
In this case, our qualitative predictions from Eq.(\ref{rde2})
should hold true. We have tested this hypothesis by performing
microscopic simulations of the birth/death process described
above. Our model consists of $N$ random walkers on a $d$-dimensional
hypercubic lattice. The walkers obey exclusion statistics.
If two walkers are at adjacent lattice sites, there is a probability
$q$ for them to spawn a new walker at a neighbouring unoccupied
site. If this occurs, a walker is randomly selected from the
entire population and is killed - thus conserving the overall
population. We have studied this model in various dimensions.
In $d=1$, and starting from a compact cluster of 100 particles,
we find that the cluster spreads out somewhat, but that the
mean cluster size asymptotes to a constant. On averaging
over realizations we find that the density profile of the 
cluster (in the center of mass frame) is in approximate
agreement with the solution of the continuum model. We
refer the reader to Fig.1 where a direct comparison is made.

In $d=2$ we find a phase transition on varying the spawning
probability $q$. For $q<q_{c}$ the cluster asymptotically
diffuses outwards, with a cluster size growing as $\sqrt {t}$.
For $q>q_{c}$ the cluster remains compact. The value of
$q_{c}$ is generally very small. For example, for $N=400$,
we find $q_{c} \simeq 0.004$. The precise $N$ dependence of
$q_{c}$, along with details of the simulations will be
given in a longer work \cite{kn}. 
At a qualitative level, we see that the continuum formulation
(\ref{rde2}) provides a good description of the underlying microscopic
model. 

\begin{figure}[htbp]
\epsfxsize=5.0cm 
\vspace*{-0.2cm}
\hspace*{1.3cm}
\epsfbox{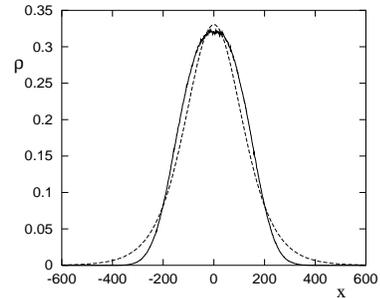}
\caption{Average (stationary) cluster density from simulations (solid line)
in $d=1$ with $N=100$ particles, compared to the continuum prediction
(dashed line) (\ref{stat2}) with $\xi _{0}$ fitted to the mean cluster size.}
\end{figure} 

In the directed polymer context there is no compelling reason
to study the system for $\lambda < 0$. However, within the
context of population dynamics, this is a perfectly
acceptable sector of parameter space to explore. In order to 
motivate such a model, we shall consider a different 
birth/death process to the one described above. Consider now
that the individuals $A$ interact with the rules:
$A \rightarrow 2A$ and 
$2A \rightarrow 0$, with rates $r$ and $r'$ respectively, which 
are in general time dependent. In other words, we have
rules corresponding to mutual annihilation and single-sex reproduction.
We choose to take $r'=r_{0}$ (constant) and tune $r$ to maintain a constant
total population. The phenomenological reaction-diffusion equation
for this process takes the form 
\begin{equation}
\label{rde3}
\partial_{t} \rho = D \nabla ^{2} \rho - r_{0}\rho ^{2} + 
r_{0}\left [ \int d^{d}r' \rho ({\bf r}',t)^{2} \right ] 
\rho \ .
\end{equation}
As desired, this corresponds to our original directed polymer 
equation (\ref{scpdep}) but with a negative coupling constant.

This equation has some remarkable properties, which may be
uncovered from an approximate analytic treatment.
We shall restrict our attention to $d=1$. We note 
that spatial coupling enters the model in two distinct ways:
locally via the Laplacian, and globally via the normalization
constraint. This means that there can be non-trivial dynamics
even if we set $D=0$ -- a limiting case which allows an exact
solution of the equation. We shall confine ourselves
to stating a few pertinent results. Details will
be given in a longer work\cite{kn}. It is obvious that in
this limit the dynamics will be extremely sensitive to
the initial condition. One static solution corresponds
to choosing the initial condition to be a normalized top-hat
function. More interesting solutions are obtained with initial
conditions with infinite support. Consider
$\rho (x,0) = (A/\xi_{0}){\rm exp}(-(|x|/\xi_{0})^{\beta })$.
For long times, one finds that this profile spreads outwards,
with a flat central region of height $\sim 1/\xi (t)$, where
$\xi (t)$ is the lateral size of the profile. 
We find that this length scale grows in time as
$\xi (t) \sim t^{\alpha }$ with
$\alpha = 1/(1+\beta )$. The solution for the density profile is
expressed as
\begin{equation}
\label{qtw}
\rho (x,t) = (1/\xi(t)) \ F \left ( (|x|/\xi _{0})^{\beta}
- (\xi (t)/\xi _{0})^{\beta } \right ) \ ,
\end{equation}
where $F(y)$ is simply related to the initial profile.
Since this solution does not have a canonical dynamical scaling
form, there may be other relevant scales; and indeed
we find that there is a second important length
scale, which is the width $W(t)$ of the interface separating 
the region of ``flat'' nonzero density and the region of vanishing
density. This quantity changes in time according to 
$W(t) \sim \xi(t) (\xi_{0}/\xi(t))^{\beta }$. Therefore 
the interface width shrinks (grows) with time for
$\beta > 1$ ($\beta < 1$). The above solution is rather
unusual as it, in some way, interpolates between a 
travelling wave solution $F(x-vt)$ and a scaling (or 
similarity) solution $F(x/\xi(t))$ \cite{baren}. We term it
a ``pseudo-travelling wave''. [We note in passing that an initial
condition with tails decaying with a power $q$ evolves under
normal dynamical scaling with $\rho \sim (1/t)F(|x|/t)$ and
$F(z) \sim z^{-q}$ for $z\gg 1$.]  

\begin{figure}[htbp]
\epsfxsize=6.0cm 
\vspace*{-0.2cm}
\hspace*{0.5cm}
\epsfbox{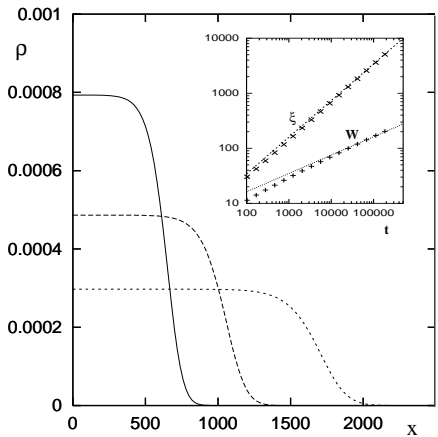}
\caption{Density profiles for three different times, 
from numerical integration of (\ref{rde3})
with $D=1$ and $r_{0}=5$. The inset shows a log-log plot of the
time evolution of the cluster size $\xi $ and the interface width
$W$. The straight lines (which are a guide to the eye) 
have slopes of 2/3 and 1/3 respectively.}
\end{figure} 

One may now ask how the above picture is modified when
one re-instates the local coupling $D$. We have several
arguments which all agree that, so long as the initial
profile has a finite transverse scale, then for late
times the density will have the form (\ref{qtw}), but
with a {\it selected} value of $\beta =1/2$. This 
implies that the profile grows outwards with a
transverse scale $\xi (t) \sim t^{2/3}$ and 
interfacial width $W(t) \sim t^{1/3}$. The simplest
(albeit non-rigorous) way to see why this value
of $\beta $ is selected is to {\it assume}
that for non-zero $D$ the density profile has
a form approximately given by (\ref{qtw}), and to then
ask how the Laplacian term in (\ref{rde3}) can
balance with the reaction terms. Since the profile
is flat for $|x| \ll \xi(t)$ and vanishing for $|x| \gg \xi(t)$
it is clear that the size of the Laplacian is set by
the width of the interfacial region. We therefore
estimate $\partial _{x}^{2}\rho \sim 1/W^{2}\xi $.
Similarly the reaction terms are of order $\rho ^{2}
\sim 1/\xi ^{2}$. Balancing the two leads us to 
$W(t) \sim \xi(t) ^{1/2}$, which on comparing with
our previous result $W(t) \sim \xi(t)^{1-\beta}$ 
gives $\beta =1/2$.
We have performed numerical integration of Eq.(\ref{rde3})
in $d=1$ for a variety of initial conditions, and with
zero or non-zero $D$. The exact results for $D=0$ have been
numerically confirmed, however the situation for
$D>0$ is less clear. In Fig.2 we show an example of the evolving
density grown from an initial Gaussian profile,
with $D=1$ and $r_{0}=5$. In the inset we plot the profile
size and interfacial width. These quantities exhibit very
strong corrections to simple power law scaling, but a 
closer analysis (studying the `running exponents') 
indicates that the asymptotic power laws are approximately
0.65(1) and 0.30(1) respectively, which are close
to the values obtained above.

In conclusion we have considered two very different
physical scenarios which share a common mathematical
description. These are self-localization of a directed
polymer, and birth/death models with fixed population
number. The physics of these processes is non-local
which leads to a number of interesting results,
including phase transitions from localized to non-localized
states, and pseudo-travelling waves. 

Finally, it is noteworthy that the fundamental directed polymer
equation (\ref{pdep}) with a general external potential $V$
can be recast in terms of constrained population
dynamics. The case of a random potential is especially interesting,
and corresponds to local, random birth/death processes, along
with a global feedback to control the population. We are currently using
this mapping to gain new insights into these two
fascinating problems.

\end{document}